\documentclass[10pt,onecolumn]{article}
\usepackage[ansinew]{inputenc}
\usepackage{dsfont}
\usepackage{enumerate}
\usepackage{varioref}
\usepackage{amsmath}
\usepackage{graphicx}
\usepackage{latexsym}
\bibliography{bibfile}
\usepackage{hyperref}

\usepackage{times}
\usepackage{mathptmx}
\usepackage{amsmath}
\usepackage{amsbsy,latexsym,bm}
\usepackage{xcolor}
\usepackage{multirow}

\def\bye{\end{document}}

\def\bye {\end{document}}

\def\beq{\begin{equation}}
\def\eeq{\end{equation}}
\def\beg{\begin{gather}}
\def\egg{\end{gather}}
\def\bla{\begin{aligned}}
\def\ena{\end{aligned}}
\def\bed{\begin{description}}
\def\led{\end{description}}

\def\bec{\begin{center}}
\def\ec{\end{center}}
\def\bet{\begin{tabular}}
\def\et{\end{tabular}}
\begin{document}

\title{\bf \large Vaccine efficacy: demystifying an epidemiological concept}
\author{\bf \normalsize Sorana Froda and Fabrice Larribe  \\
 \bf \normalsize D\'epartement de math\'ematiques, C.P. 8888, UQAM, Montr\'eal, Qu\'ebec, Canada}
\maketitle

\vskip20pt

\noindent {{\bf Abstract.} The paper is an introduction to the concept of vaccine efficacy and its practical applications. We provide illustrative examples aimed at a wider audience.}


\vskip20pt

\section*{\large 1. Introduction}
{During the recent} Covid-19 pandemic \cite{WHO-covid} various epidemiological terms have started to widely circulate in the media and were often cited by decision makers. Moreover, the pressure has increased in informing the public about numerical values which are hard to comprehend without a more precise context. Thus, the general public has started to deem as good or bad some epidemiological measures, in particular vaccine efficacy, by some vague standards, and feel critical if some specific {value} is not met, in spite of assurances given by health authorities. 
This being said, this type of concept has a probabilistic or statistical interpretation which is hard to be made precise in every day language. 
Indeed, what is meant by 70\% efficacy when a vaccine is concerned? To our na\"{\i}ve ears this value suggests that 30\% of the time one is not actually covered by the vaccine and could get sick, but what does this mean more precisely? 
Thus, it may be worth trying to clarify the concept in layman's terms.

In the present paper we consider the temporal aspect of vaccine immunity, and translate into layman's terms some statistical aspects of the recent studies on Covid-19 vaccines. To put it roughly, in the Moderna \cite{moderna}, Pfizer \cite{pfizer}, or AstraZeneca \cite{astra-z} Phase-3 vaccine developments, the approach was similar: there was a follow-up of two large cohorts of individuals of similar characteristics (age and some physical or medical conditions considered to be risk factors for a more serious illness, like obesity). At the starting time which was not the same calendar time for all participants,  the subjects in one cohort were injected the new vaccine, the other ones a placebo (a substitute with no impact on immunity), based on a random assignment of subjects to vaccine and placebo.
And then, the researchers  waited $\dots$ Well, actually, they periodically tested the individuals in the cohorts, assessed symptoms, their severity, the participants' health condition, etc. Although  the whole process was more complicated than pictured here, and each study performed slightly different assessments, this summary description suffices for the main purpose of the present discussion. Last point: this type of study is called a {\it randomized controlled trial}, or RCT.

\section*{\large 2. Vaccine efficacy}

\subsection*{\large 2.1 The concept and numerical examples}

In order to illustrate the concept of {\it vaccine efficacy} (VE), 
consider simple hypothetical numerical values. \footnote {Indeed, the examples used in this paper are illustrations that make calculations straightforward. We do not reproduce data and results published in specialized journals where the papers can present more sophisticated analyses. Still, the order of magnitude is comparable with what can be encountered in practice.
}
Assume there are  two studies, one for vaccine~A and another one for vaccine~B. Further, in each study, there are two cohorts (vaccinated and non vaccinated) each with 10,000 participants that are followed up over time. Moreover, assume that we stop 20 days after injecting  the vaccine 
and count those who tested positive for the disease. Further, it is reported that Vaccine A has vaccine efficacy ${\rm VE_A}$=60\% and  vaccine B 
has ${\rm VE_B}$=80\% (we assume that immunity takes place immediately after inoculation, for simplicity). How was this calculated, i.e. how many people really fell ill (e.g. tested positive) in each cohort? Let's designate  such an {\it infected} person by the current medical terminology, namely {\it a~case}.

In what follows, we assume that during the time of the study we have, { in the general population},  {\it an infection rate} of one case among ten thousand per day (which is a realistic rate for some periods of the present pandemic). So, in a  cohort  of ten thousand we have on {\it average} (or we {\it expect}) one infection a day and after 20 days there should be 20 cases among the non vaccinated people in each study, A or B. With the vaccine,  
one expects to have less cases among the vaccinated and the vaccine efficacies indicate how to measure this reduction in cases. With ${\rm VE_A}$=60\% we have 60\% less cases, while with ${\rm VE_B}$=80\% we have 80\% less cases; then, in our example the {\it expected} number of cases are given by: 

\begin{figure}[!ht]
\begin{center}
\includegraphics[scale=0.3]{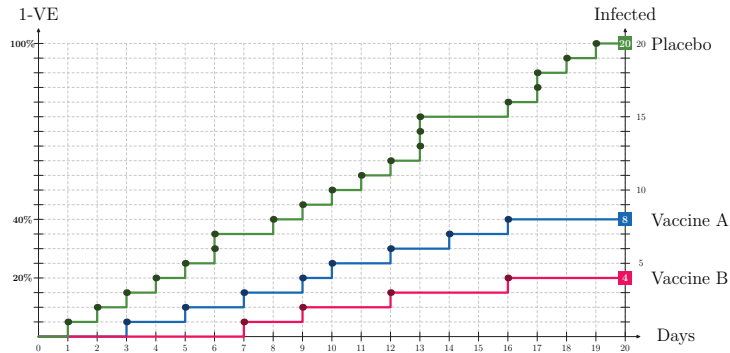}
\end{center}
\caption{\small Two vaccines and a placebo, an illustration of arrivals of cases over time, where the daily incidence rate per 10,000 is 1, ${\rm VE_A}$=60\%, and ${\rm VE_B}$=80\%; we count the cases during 20 days after inoculation; the height of every step gives the number of new cases in a specific day; in the graph we have 
0, 1, 2, or 3 new cases in any given day.}
\end{figure}

\begin{center}
Study A: 40\% = (100-60)\% out of 20 or 8 cases among the vaccinated, \\ 
Study B: 20\% = (100-80)\% out of 20 or  4 cases among the vaccinated.
\end{center}

So, after 20 days,  in study A we should have 8  cases out of 10,000 participants, while in Study B we should have half as many, 4 cases out of 10,000. So, we can see that these are very few cases, even in Study~A with the  {poorer} quality vaccine there are less than one in a thousand cumulated cases among the vaccinated.

Further, let's take a closer look  at our calculations. First, in medical studies, calculations are done {backwards}, from actual number of cases to percentages, 
since we try to evaluate (in specialized language {\it estimate}) VE, which is unknown. So, cases are counted in each cohort and VE is assessed (estimated) based on the ratio of the counts. So, in order to get an estimate of the VE one would compute, after 20 days (given our numerical example):
\eject \vfill

\begin{center}
Study A: 8/20=0.4=40\% and ${\rm VE_A}$=1-0.4=0.6=60\%,\\
Study B: 4/20=0.2=20\% and ${\rm VE_B}$=1-0.2=0.8=80\%.
\end{center}
Of course, computing 8/20 (or 4/20) is equivalent  to computing a ratio of  proportions
\bec
\bet{l}
8/20=(8/10,000) / (20/10,000),  \\ 
 4/20=(4/10,000) / (20/10,000),
 \et
\ec 
given that in both cohorts (vaccinated and non vaccinated) we have the same number of participants. 
Another way of getting the ${\rm VE_A}$, ${\rm VE_B}$ values given above is to 
compute directly the reduction in cases due to vaccination; in Study~A we go down from 20 to 8, in Study~B, we go down from 20 to 4, and we obtain the same percentages:
\begin{center}
\bet{l}
Study A: ${\rm VE_A}$=(20-8)/20=12/20=0.6=60\% ,\\
Study B: ${\rm VE_B}$=(20-4)/20=16/20=0.8=80\%.
\et
\end{center}

To summarize, VE quantifies the {\it reduction in risk} of getting {\it infected} where one compares a vaccinated with a non vaccinated person; thus, if VE=80\% this risk is reduced by 80\%, i.e. we get 80\% less cases among vaccinated, or, among vaccinated we have a  fifth (=20\%) cases as among non vaccinated. 
\footnote {Infection is a term used loosely here, as it could mean mild or severe symptoms, etc.}
Finally, consider two extreme situations: (i) VE=0\% where among vaccinated the risk is the same as among non vaccinated, the vaccine has no effect; (ii) VE=100\% where there is no infected among the vaccinated. The term {\it risk} can itself correspond to more than one measure, but in this article we use mainly the concept that takes into account the time spent in the study  (we count cases per day, among 10,000); our computing is similar to the assessments \footnote {Indeed, in principle we compare proportions of new cases per time unit but there is no difference with comparing proportions of cumulated cases when
the follow-up time is the same for all participants. 
The specialized literature includes a correction to take into account the variations in the follow-up time.}
done in \cite{pfizer},\cite{moderna}, or~\cite{astra-z}.  

In other words, 
 VE {\it compares the proportion of  cases among vaccinated and non vaccinated}, and the  infected among the non  vaccinated may be very, very few to start with 
and their proportion in the observed cohort very small.  This proportion of infected has nothing to do with the percentage expressed in the VE value. To put it otherwise, VE is a {\it relative measure}, not a population percentage, not a cohort percentage. 

Further, it is worth noting that in a vaccine study like the ones cited above, the researchers can have a full portrait on how cases are registered over time and are able to perform more refined analyses. \footnote {\label{time} For example, every time a case is notified, the total of initially 10,000 non diseased participants is reduced, but in our examples these intermediate values until day 20 are so close to 10,000 that in our calculations we have ignored them.  This will be our convention for the rest of the paper but it's important to note that time plays an important part in all these assessments.} 
This being said, our argument makes it clear that the VE computed in any vaccine randomized controlled study (RCT) is more or less a snapshot in time, during the weeks where the study is conducted. Even if the VE stays more or less stable over long lapses of time, the actual proportions, like the ones given above, of 8 or 4 in 10,000 can change along the progress of the infection in the population, given its varying {\it incidence} (proportion of new non vaccinated cases per time unit).

\subsection*{\large 2.2 Taking into account heterogeneity}

We can refine the above calculations since, as mentioned above, other variables are measured in such a study, in particular age. So let's imagine that we have again the same risk of one in 10,000 per day of getting infected but the vaccine acts differently on two age groups,
{\it old} and {\it young}, where: VE({\it old})=50\% and VE({\it young})=70\%.  Assume that we follow up two cohorts of 20,000  each (one vaccinated, the other one not), where half  are {\it old} and the other half are {\it young}. Then,  
among the non vaccinated we {\it expect} to get 2 infected per day and 40 infected after 20 days, split into 20 {\it old} and 20  {\it young}. As for the vaccinated,
by the same type of computing as above we end up with 10 cases among the {\it old} 
and 6~cases among the {\it young}, since
\bec
20$\times$50\%=10\,,\;20$\times$30\%=6\,.
\ec
 This situation is summarized in the following table.
\bec
Situation~1:  follow-up 20 days; half {\it old}, half {\it young}
\medskip

\bet{|c|c|c|c|c|c}
\hline
  &   \multicolumn{2}{|c|} {Non vaccinated } &  \multicolumn{2}{|c|} {Vaccinated}    \\
 \cline{2-5} 
& Total & Infected  & Total & Infected \\
\hline
Old & 10,000 & 20 & 10,000 & 10  \\
Young & 10,000  & 20 & 10,000 &  \;6\\
\hline
Total & 20,000& 40& 20,000 & 16 \\
\hline
\et
\ec 
If we try to get a unique VE value by combining the two age groups,  we can simply add the vaccinated cases in both groups, as well as the non vaccinated cases in both groups, and compute the ratio of sums, namely
\bec
(10+6)/(20+20)=16/40=0.4,
\ec
 which gives VE=1-0.4=0.6=60\%.
We notice that 60\% is nothing else but the mean of VE({\it old}) and VE({\it young}), i.e.
\bec
  [VE({\it old})+VE({\it young})]/2=[50\%+70\%]/2=60\%.
\ec
So, in a heterogenous group, the VE value is a  compromise of the vaccine success in the two subgroups, as 
$50\%<60\%<70\%$.

To go a bit deeper in this calculation, we can imagine a  more complicated setting, where we would have less  {\it old} participants than {\it young} participants, which is a bit reflective of what happens in real life. For simplicity, assume that the two cohorts, of 20,000 each, are again divided in {\it old} and {\it young}, but with
three times more  {\it young} than  {\it old} participants. This would give  5,000 {\it old} and 15,000 {\it young} participants in each cohort (one cohort vaccinated, the other one non vaccinated), or one fourth  {\it  old}, three fourths  {\it young}. 
What happens in this second situation  
is summarized in the following table.
  \bec
Situation~2: follow-up 20 days; three times more {\it young} than {\it old}
\medskip

\bet{|c|c|c|c|c|c}
\hline
  &   \multicolumn{2}{|c|} {Non vaccinated } &  \multicolumn{2}{|c|} {Vaccinated}    \\
 \cline{2-5} 
& Total & Infected & Total & Infected \\
\hline
Old & \; 5,000 & 10 & \; 5,000 & 5  \\
Young & 15,000  & 30 & 15,000 &  9\\
\hline
Total & 20,000& 40& 20,000 & 14 \\
\hline
\et
\ec
We have again 40 cases (infected) among the non vaccinated but they are divided unequally among {\it old} and  {\it young} and  in the same proportion (1 to 3) as in the cohort, namely there are 10 {\it old} and 30 {\it young} infected. Further, we
compute as before, and apply to each age class the  appropriate percentage (100-VE)\% in order to get the number of cases among the vaccinated: 
\bec
\bet{l}
 10 $\times$ 50\%   = 5 cases among the {\it old} vaccinated,\\
30 $\times$ 30\%  =  9  cases among the {\it young} vaccinated.
\et
\ec

Thus, this time, the combined vaccine efficacy is
\bec
1-(5+9)/(10+30)=1-14/40=1-0.35=0.65=65\%,
\ec
again a value in-between VE({\it old}) and VE({\it young}) but higher than before and closer to VE({\it young}) as the {\it young} are more numerous. 

Thus, in order to get a general formula for the efficacy of the combined group, we have to take into account this unequal division among the two age groups and get a weighted mean of VE({\it young}) and VE({\it old}); namely, in our numerical example with 0.25=25\% {\it old}, 0.75=75\% {\it young} the combined efficacy computed above is given by the weighted formula:
\bec
VE = VE({\it old}) $\times$ 0.25 + VE({\it young}) $\times$ 0.75=
{0.5} $\times$ 0.25 + {0.7} $\times$ 0.75 
=0.65
= 65\%.
\ec
This example illustrates the fact that the final VE value can depend on such proportions of subgroups and justifies the need (in many studies) to control and take into account the age distribution or health status of the participants or any other variable that could have an impact on the efficacy of the vaccine.  
We can also see that the final result would be different if 
the weights changed, e.g. with 0.15=15\% {\it old} and 0.85=85\% {\it young}, we get a higher value, as expected:
\bec
VE = VE({\it old}) $\times$ 0.15 + VE({\it young}) $\times$ 0.85=
0.5 $\times$ 0.15 + 0.7 $\times$ 0.85=0.67
=67\%.
\ec

\subsection*{\large 2.3 From relative risks to odds ratios}

Before proceeding, we briefly describe an alternative formula, that will allow us to consider another measure of the quality of a vaccine. We consider again vaccine~B as above, but rather than comparing the proportions 8/10000 and 20/10000 we can compare the {\it odds}, that express  how the 10,000 participants in each cohort are divided among infected and non infected. The odds are: 
\bec
\bet{l}
ODDS (infection given vaccine)=4/(10000-4),
\\
 ODDS (infection given placebo)=20/(10000-20), 
\et
\ec
 and their ratio is
\beq
\text {OR}={4/(10000-4) \over 20/(10000-20)} \approx {4/10000 \over 20/10000} = 20\%.
\eeq
 Roughly, the odds of falling ill among vaccinated is only a fifth of the odds of falling ill among the non vaccinated. The actual value of OR  is 0.1996799.
 In this situation the ratio of the {\it odds} is almost the same as the ratio of proportions 
 because the proportion of cases (infected)  is  very 
 small, a consequence of the fact that the incidence rate is one in 10,000 per day;
 thus, in this particular setting, VE$\approx$1-OR. This concept, i.e. OR, allows us to make the connection with an equivalent quality measure, 
 the {\it  vaccine effectiveness}, whose description is left to the Appendix.

\section*{\large 3. Vaccination policies: a comparison}

Finally, we address another issue, 
of great concern to the public: vaccination policies. We propose to compare two vaccination policies, postponing or not the distribution of a second dose of a vaccine during a specific time window, 
under the following assumptions:
\bec
\begin{tabular}{l}
--non vaccinated infection rate of one in ten thousand;\\
--VE1=60\% as the efficacy  
after the first dose (once injected);\\
--VE2=80\% as the efficacy  
after the second dose (once injected);\\
--a cohort of 20 thousand people for each vaccine.\\
\end{tabular}
\ec
The vaccines proposed in \cite{pfizer}, \cite{moderna}, \cite{astra-z} have no immediate immunization effect but to ease the calculations we have proposed to consider a less complicated setting where there is no immunity lag after  inoculation. 
Thus, in order to illustrate our comparison, we consider two cohorts of twenty thousand each, divided in two halves, H1 and H2,  a follow-up time of 40 days divided in two time slots of 20 days each, and two very simple scenarios. 
In both scenarios the same number of people are vaccinated in each time slot.
\bec
\bet{l}
Scenario~1. Half the cohort 
receives two doses, half is not vaccinated by day 40: 
\\
\bet{l}
H1 receives dose 1 (day 1), dose 2 (day 21),\\
H2 is not yet vaccinated by day 40.\\
\et
\\ \\
Scenario~2. Everybody gets one dose by day 40, as follows: \\
\bet{l}
H1 receives dose 1 on day 1 \\
H2 receives dose 1 20 days later (day 21).\\
\et
\et
\ec
What scenario is the winner? In other words, what scenario ends up with less infected (cases) after 40 days? Let's divide the time in 20 days time slots, and count how many infections (cases) can occur during the first and last 20 days, among the H1 or H2. In the first scenario, H2 is not vaccinated and after 40 days there are 40 cases because we asume one infection per day with no vaccine; in the second scenario H1 gets dose~1 on day one, so its infection rate (of one in ten thousand per day) is reduced by 60\%, and by day 40 only 40$\times$40\%=16 get infected. 
The second scenario appears to be the winner, as intuition would have told us.  The full computing can be
 a bit elaborate, because there is progress over time, and in each scenario, once some people get infected they can no longer contribute new cases in the future. Still we can ignore how the 10,000 are reduced over time, given the very small numbers of infected \textsuperscript{\ref{time}}. We end up  
with  the (approximate) tables below, computed as follows:  
at the end of each time slot, 
the expected number of cases with no vaccine would be 20 cases; further, according to the vaccination scenario, we apply the apropriate percentage to the 20 cases in each 20 days time slot,  namely 100\%, 40\% or 20\% ( i.e. no vaccine, after dose~1, after dose~2).


\bec
Scenario~1: 52 cases in total 
\medskip

\bet{|c|r|r|c|c|c|}
\hline
&   \multicolumn{2}{|c|} {Vaccinated in each cohort} &  \multicolumn{2}{|c|} {New cases}   & Total cases  \\
 \cline{2-6} 
Cohort   & Day 1: dose 1 & Day 21:dose 2 & Days: 1-20 & Days: 21-40& Day 40\\
 \hline
H1& 10,000 & 10,000 & \;8 & \;4 & 12\\
H2&   0 & 0 & 20 & 20 &40\\
\hline
Total & 10,000& 10,000 & 28 & 24 & 52\\
\hline
\et
\ec

\bec
Scenario~2: 44 cases in total\;
\medskip

\bet{|c|r|r|c|c|c|}
\hline
 &   \multicolumn{2}{|c|} {Vaccinated in each cohort} &  \multicolumn{2}{|c|} {New cases}   & Total cases  \\
 \cline{2-6}
Cohort  & Day 1: dose 1 & Day 21: dose 1 & Days: 1-20 & Days: 21-40 & Day 40\\
 \hline
H1 & 10,000 & 0 & \;8  & 8 & 16\\
H2 & 0 &  10,000 &  20 & 8 & 28\\
\hline
Total   & 10,000 & 10,000 & 28 & 16 & 44\\
\hline
\et
\ec

\begin{figure}[!ht]
\begin{center}
\label{scenarios}
\includegraphics[scale=0.085]{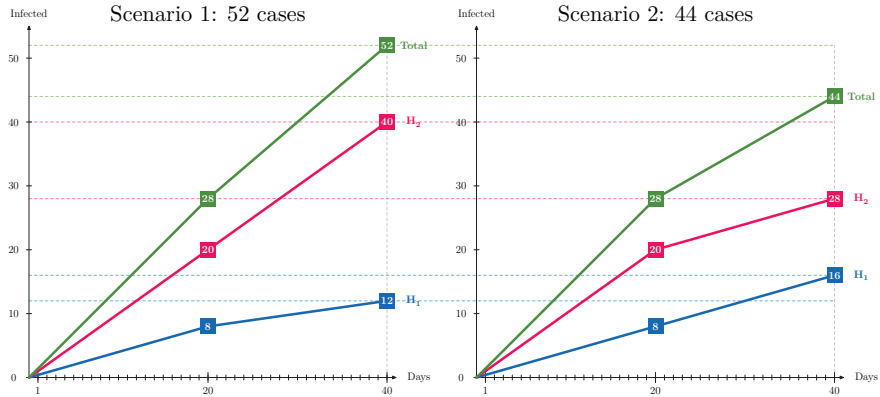}
\end{center}
\caption{\small Expected number of infected after 40 days, in two vaccination scenarios with 10 000 jabs at day~1 and day~21.}
\end{figure}

We notice that the big difference between scenarios (after 40 days, in Scenario~1 we have 18\% more cases) comes from the fact that in Scenario~1 half the people keep getting infected at the non reduced rate as only part of the population is vaccinated, albeit with two doses. So, this small example suggests that to postpone the second dose and continue vaccinating with dose~1 may not be a bad idea  (statistically). 
This being said, in the real-world many other considerations are at stake, for instance severity, mortality, immunity coverage after one dose, etc. Indeed, if all people in H1 are more prone to severe forms of the disease, during the second time slot the number of such serious cases  is doubled in Scenario~2 as compared with Scenario~1 and this can be problematic. \footnote {Just multiply by 100 these numbers i.e. consider cases per million; then, by choosing Scenario~2 over Scenario~1, we have 800 instead of 400 cases among H1, during days 21-40.}
On the other hand, it makes sense to think that H2 has many more people than H1, e.g. there are twice as many individuals in H2 than in H1. Then the difference between total infected is even bigger as in both scenarios we double the number of infected that originate from H2 and we get, in the end: 
\bec
12+80=92 infected (Scenario~1) versus 16+56=72 infected (Scenario~2), 
\ec
i.e. there are 27\% more cases in Scenario~1 than in Scenario~2. In  this last setup, Scenario~2 doubles the number of problematic cases, while Scenario~1 increases by more than 25\% the total number of cases. Moreover, Scenario~2 that ends up with less cases requires more vaccinations than Scenario~1. Hence,  there is no clear winner and we realize that finding a  best plan is not straightforward. Indeed, we see that decisions must take into account many aspects of a specific population, like age distribution, as well as a complex field reality, for instance: using more than one vaccine brand where each one has a different VE, scheduling injections over all days of the  week (not at fixed days like in our scenarios),  taking into account immunity lag,~etc. Definitely, it is not for us to advise health agencies, which, by the way, use much more elaborate models than the layman's examples given~here.

\section*{\large 4. Concluding remarks}

To recapitulate, is VE=50\% good or bad? What we know is that it {\it reduces} by 50\%  the risk of some deleterious event; generically, let's call this event a {\it new infection}, where, depending on the context, this {new infection}
can mean getting
a mild or a severe form of the disease, only a symptomatic or both a symptomatic and an asymptomatic condition, being hospitalized, etc. Therefore, it is not uncommon to see reports of different efficacies for the same vaccine, according to the type of event considered. Typically, with no vaccine, the risk of getting infected can still be very small, like one new case per day among 10,000. In this example,  a vaccine with VE=50\% would reduce the risk to one case  per day, among 20,000. If we consider that originally we could expect to have 20 infected out of 10,000 after 20 days, a vaccine with a VE=50\%  reduces this expected number to 10 infected (among 10,000) after 20 days.  A high VE (like 80 or 90 percent) reduces to a trickle the risk of getting infected (one in 50,000 or one in 100,000 per day, respectively). Or, to put it differently,  after 20 days, rather than having 20 infected among 10,000 we end up with only 4, respectively~2, infected. In a population of one million, 2,000 is reduced to 400, respectively 200, and the order of magnitude is changed.

There are other ways of looking at these {\it risks} (or {\it incidence rates}): rather than expressing the risk by  referring to the number of cases per 10,000 per day, imagine we follow up 100 people for 100 days. Then, the incidence rate of one in 10,000 per day 
comes to 
saying that one of the 100 people could get infected after 100 days; consequently, 2 out of 100 people could get infected after 200 days. Thus, with a vaccine efficacy of 50\% we expect to have only one infected among the 100 after 200 days (or more than 6 months). So,  in a time window of more than 6 months a VE=50\% reduces the percentage of infected  in half, and this would give an important head start in fighting the disease. This being said, the counts in this article are only illustrations and the numbers are estimates that depend on the incidence rate with no vaccine, i.e. something like an {\it affliction propensity}, which varies over time and by location. This {\it affliction propensity} can be higher or lower than one in 10,000  per day and by CDC standards  for the Covid-19 this is a very high incidence \cite {CDC-travel}.
Given these fluctuations, the term {\it risk} must be taken with a grain of salt as well, as it can contain a reference to time (getting infected after two weeks, e.g.). Also, references to time comprise a starting point, and only in a controlled (RCT) study time is well measured from the moment participants are given a vaccine dose, or a placebo, or are enrolled,~etc.

To summarize, the {\it vaccine efficacy} (VE) is a relative measure, and reflects the reduction in affliction propensity 
where vaccinated are compared with non vaccinated.
This being said, many more factors enter into consideration, like age distribution in a specific population and severity of the disease. Severity and hospitalizations play a crucial role in the case of the Covid, and even a small affliction propensity can lead to catastrophic results.

 So, what about  a {\it vaccine efficacy} of 50\%, is it big enough? Well, one should feel reassured: the 50\% is a {\it reduction} in the risk of being ill and not the {\it actual} risk that is typically very low;
with a VE=50\% the vaccination cuts this risk in half.
 With present day Covid vaccines all efficacies are higher and the reduction in risk is much more important.
 \footnote {For example, in the placebo arm (non vaccinated cohort) of the Pfizer-BioNTech study \cite{pfizer} the incidence (risk) was less than 2 in ten thousand per day. Their reported efficacy after two doses is VE=95\%. In the Astra Zeneca early study \cite{astra-z} the placebo arm had an incidence of 4 in ten thousand per day, and their reported overall efficacy is VE=62\%. Finally, in the Moderna study \cite{moderna}, the per-protocol  incidence was 1.5 in ten thousand per day, and the reported efficacy after two doses is VE=94.1\%.
}
Moreover, as far as severe cases go, most vaccines have an efficacy close to 100\%, since there were no severe cases among the vaccinated. This being said, even a vaccine with a VE=50\% would reduce in 
half severe cases or deaths, an improvement that no society can afford to take lightly.

\parindent=10pt

\normalsize

\vskip10pt
\eject \vfill
\bec
{\bf \large Appendix A:}
{\large From vaccine {\it  efficacy} to vaccine {\it   effectiveness}}
\ec
This section is a short digression into a practical issue: how can we continue assessing (or confirming) VE under real-world conditions, once  the controlled study has  been completed? In a real life context, we do not deal with cohorts that we follow up {\it forward} in time ({\it prospective} approach), one vaccinated and one that is not vaccinated. On the other hand, it is of great interest to assess the success of a vaccine but then  we have to resort to a different method.  Although very important in practice, we give only a quick outline of this alternative method, for the curious  (as the technical part can become too burdensome). 

In such situations, one makes a comparison {\it backward} in time ({\it retrospective} approach):  we start with two groups, one infected, the other one not infected, and  we count how many in each group have been previously vaccinated or not. In other words, we proceed the other way round: we know who is infected and who is not, and we check their vaccination status, with two advantages: (i) we do not have to wait for the infection to take place, as cases  are available at the time of the study; (ii)~we base the analysis on a larger number of infected,
a very useful tool when we deal with a rare disease: for instance, we can study as many infected as non infected although there are only 1 in 10,000 infected.
On the other hand, we can no longer control the age and other characteristics of the participants; this is an {\it observational case-control} study. 
Moreover, there is no reference to time and we end up with a different but similar measure to VE, the so-called {\it vaccine effectiveness}, VVE. It is worth mentioning that vaccine effectiveness is assessed yearly by the CDC for the seasonal flu \cite{CDC-flu}. 
To illustrate the concept, assume we sample 180 infected and 180 not infected, and each group has the following split: 
90 non infected have been vaccinated, while only 30 infected have been vaccinated; 
the odds of vaccination are: 
\bec
\bet{c}
ODDS (vaccine in infected)=30/150=1/5,\\ i.e. 1 vaccinated to 5 non vaccinated, \\
ODDS (vaccine in non infected)=90/90=1, \\i.e. 1 vaccinated to 1 non vaccinated.\\
 \et
\ec
Such  data is usually presented in a cross  
 table, by counting the number in each of the four classes and their totals:
\bec
\bet{|c|c|c|c|c|}
\hline
& \bf Total & Vaccine & No vaccine  & Odds\\
\hline
\bf Total  & \bf 360 & \bf 120 & \bf 240   & \\
\hline
Infected  & \bf 180&  30 & 150 & 30/150=1/5\\
\hline
Not infected & \bf 180 & 90  & 90  & 90/90=1\\
\hline
Odds &  & 30/90=1/3 & 150/90=5/3 &  OR=1/5\\
\hline
\et
\ec
If we compare  the odds of vaccination among infected and non infected we get 1/5 versus 1, and the odds ratio is OR=1/5. This OR value gives a sense to what extent there are less vaccinated people among the infected than among the non infected. 
Therefore, by definition, the {\it vaccine effectiveness} VVE=1-OR=1-0.2=0.8=80\%, and VVE is also a relative measure. Indeed, the same VVE value could be obtained with many other splits among vaccinated and non vaccinated, e.g.: 
\eject \vfill
\bec
1/4, i.e. 1 vaccinated to 4 non vaccinated (infected) and \\
5/4, i.e. 5 vaccinated to 4 non vaccinated (non infected)\\
 or \\
 1/1, i.e. 1 vaccinated to 1 non vaccinated  (infected) and \\
 5/1, i.e. 5 vaccinated to 1 non vaccinated (non infected).
  \ec
\footnote {Actual numbers of vaccinated/non vaccinated: 36/144 (infected) and 100/80 (not infected); 90/90 (infected) and 150/30 (not infected).} 
This being said, we can also consider the odds of infection among vaccinated and non vaccinated that are, respectively: 
\bec
1/3, i.e. 1 infected to 3 non infected (vaccinated) and \\
5/3, i.e  5 infected to 3 non infected (non vaccinated)
\ec
and their ratio OR=1/5, as before.
The   ratio of the odds of infection 
among vaccinated and non vaccinated  turns out to be the same as the ratio of the odds for being vaccinated computed by selecting infected and non infected. 

This development gives us the tools to connect  {\it vaccine effectiveness} to {\it vaccine efficacy}. For the example seen in Section~2 we noted that  the ratio of the odds  of infection
in vaccinated and non vaccinated comes, approximately, to the ratio RR of the risks of getting infected  
 in each category. 
 As, by definition, VE=1-RR, we have also VE$\approx$1-OR=VVE, as illustrated numerically in Section~2. Well, there is a catch in this  equality: the approximation does not always work,  
  it works only if the proportion of infected is very very small, as in the example considered in Section~2, where we
followed up a large cohort over time, and the rate of infection was one in 10,000 per day.~\footnote {For example, in the table above, the proportions of infected are 1/4 for vaccinated and 5/8  for non vaccinated; thus, the ratio of proportions is  
  RR=40\% the double of  OR=20\%.}
This being said, in the general population this  is what happens: quite few people get infected over time, whether vaccinated or not vaccinated. As we have no direct way of assessing the value RR directly unless we follow up a large cohort over longer periods of time we can approximate RR by using
a ``clever trick'', and compute the ratio of the odds, after counting those who got the vaccine among infected and non infected.
This kind of effectiveness study was conducted in England to assess the quality of the Pfizer-BioNTech and AstraZeneca vaccines, and the results can be found in a submitted paper~\cite{two-vaccines}.

On the other hand, during the present Covid-19 pandemic, health agencies have also compared new cases among those who have been vaccinated  and those who haven't been, in two comparable cohorts (for example health workers) during a specific time window, a method  that  is similar to the controlled study based on two cohorts described above; see, for instance, \cite{CDC-pfizer-moderna},~\cite{israel}. Actually, the CDC has a comprehensive program of conducting various studies and applying more than one method in order to continue an asessment of the quality of vaccines and success of the vaccination campaign, as listed in \cite{CDC-covid-vaccine}.

\eject \vfill

\bec
{\bf \large Appendix B:}
{\large The far side of the Moon: the hidden world over there}
\ec
As stated in the Introduction, this article is an attempt to translate some epidemiological and statistical concepts into layman's terms. In order to achieve this we presented over simplified numerical scenarios and calculations and we did not explicitly refer to underlying specialized concepts in statistics either. 
For instance, there are techniques in the statistical analysis  of contingency tables, survival analysis, Poisson regression or Poisson processes that are behind 
the assessment of the quality of the vaccines (see, e.g. the Principles of Epidemiology  \cite{CDC-book} produced by the CDC, or a plethora of  textbooks in biostatistics and survival analysis). Moreover, although we  preserved the order of magnitude, we used illustrative simple values that do not correspond to any vaccine study in particular.
 
 Finally, one major technical aspect that we did not dwell into involves the uncertainty inherent in any probabilistic computation. This is expressed in the  {\it confidence intervals} 
that are reported in papers like \cite{pfizer}, \cite{moderna}, \cite{astra-z}, \cite{israel} or \cite{two-vaccines}. The main idea is that all percentages presented in these studies were obtained with their data in hand but they estimate a population value, fixed and unknown, a {\it parameter}, that does not depend on the subjects in the study: the parameter can be viewed as  an {\it ideal} or  {\it grand average} value.
Therefore, the parameter is not (exactly) equal to the value obtained from one data set but this data value   
{estimates} (approximates) the unknown parameter.  Moreover, based on the same data set we can give a probabilistic guess how large or small this parameter could be: these are the lower and upper bounds of the reported confidence interval. The concept of confidence  interval goes well beyond the scope of the present discussion but we mention it for two reasons: (i)~ it~is an indirect measure on the precision of the approximation, as quantified by the length of the interval; (ii)~the {\it exact} efficacy (or effectiveness) should lie in-between the bounds of these intervals, and therefore can be higher or lower than the data value,  a fact worth mentioning. 

\end{document}